\documentclass[
 amsmath,
 amssymb,
 aps,
 reprint,
 superscriptaddress,
 nofootinbib,
 floatfix,
]{revtex4-1} 

\usepackage[colorlinks=true,linkcolor=blue,citecolor=blue,urlcolor=blue]{hyperref}
\usepackage[capitalize]{cleveref}
\usepackage{bbm}
\usepackage{bm}
\usepackage{dsfont}
\usepackage{graphicx}
\usepackage{mathtools}
\usepackage{subcaption}
\usepackage{tikz}
\usepackage{xcolor}
\usepackage{ulem}
\usepackage{appendix}

\DeclarePairedDelimiter\bra{\langle}{\rvert}
\DeclarePairedDelimiter\ket{\lvert}{\rangle}
\DeclarePairedDelimiterX\braket[2]{\langle}{\rangle}{#1 \delimsize\vert#2}

\DeclarePairedDelimiter\abs{\lvert}{\rvert}

\renewcommand{\vec}[1]{\bm{#1}}

\begin{document}

\title{Quantum Monte Carlo calculations of dark matter scattering off light nuclei}

\author{Lorenzo Andreoli}
\affiliation{Theoretical Division, Los Alamos National Laboratory, Los Alamos, New Mexico 87545, USA}
\affiliation{Dipartimento di Fisica, University of Trento, via Sommarive 14, I-38123 Povo, Trento, Italy}
\affiliation{INFN-TIFPA, Trento Institute for Fundamental Physics and Applications, 38123 Trento, Italy}

\author{Vincenzo Cirigliano}
\affiliation{Theoretical Division, Los Alamos National Laboratory, Los Alamos, New Mexico 87545, USA}

\author{Stefano Gandolfi}
\affiliation{Theoretical Division, Los Alamos National Laboratory, Los Alamos, New Mexico 87545, USA}

\author{Francesco Pederiva}
\affiliation{Dipartimento di Fisica, University of Trento, via Sommarive 14, I-38123 Povo, Trento, Italy}
\affiliation{INFN-TIFPA, Trento Institute for Fundamental Physics and Applications, 38123 Trento, Italy}

\date{\today}

\begin{abstract}
We compute the matrix elements for elastic scattering of
dark matter (DM) particles  off light nuclei
($^2$H, $^3$H, $^3$He, $^4$He and $^6$Li) using quantum Monte Carlo methods.
We focus on scalar-mediated DM-nucleus  interactions
and use  scalar currents obtained to next-to-leading order in chiral effective theory.
The nuclear ground states are obtained from a phenomenological nuclear
Hamiltonian that includes the Argonne $v_{18}$ two-body interaction and
the three-body Urbana IX interaction.
Within this  approach, we study the impact of one- and two-body currents and  discuss the size of nuclear uncertainties,
including for the first time  two-body effects in $A=4$ and $A=6$ systems.
Our results  provide the  nuclear structure input needed  to assess the  sensitivity of future
experimental searches of (light) dark matter using light nuclei,  such as  $^3$He and $^4$He.

 \end{abstract}

\maketitle

\section{\label{sec:Introduction}Introduction}

Observational evidence for dark matter (DM) in the universe is extremely
strong, coming from both astrophysics and cosmology~\cite{Bertone:2016nfn}.
While searches for signals from direct, indirect,  and accelerator experiments have
yet to be successful,  a vibrant  worldwide experimental program exists.
In particular,  the so-called ``direct detection'' search for weakly interacting massive particles
through nuclear recoils is very active,
and there is a growing emphasis on covering a broader DM mass range,
extending to the sub-GeV scale~\cite{Battaglieri:2017aum}.

As  emphasized already in early studies~\cite{Engel:1992bf},
to interpret  direct detection experiments and  disentangle the origin of possible future signals,
it is important to have a solid  theoretical control of nuclear effects.
In recent years,  a variety of  approaches based on  effective field theory (EFT)
have been proposed to tackle the physics of DM-nucleus interactions.
EFT methods have been applied at different levels:
(i) nonrelativistic DM-nucleus interactions~\cite{Fan:2010gt};
(ii)  nonrelativistic DM-nucleon interactions~\cite{Fitzpatrick:2012ix};
and (iii)  DM-nucleon interactions derived from DM-quark and DM-gluon interactions
in the framework of chiral EFT~\cite{Prezeau:2003sv,Cirigliano2012,Menendez:2012tm,Hoferichter:2015ipa,Hoferichter2016,Korber2017,Bishara:2017pfq},
to be  used in nuclear few- and  many-body calculations.
First-principles, lattice QCD calculations, have also been performed for matrix elements of scalar, axial, and tensor currents~\cite{PhysRevLett.120.152002,Beane:2013kca}.

We work within approach (iii),  which is  the only one suitable for matching to higher scales and performing a consistent
phenomenology of direct, indirect, and collider DM searches.
In this approach,  several classes of operators arise at the DM-quark and DM-gluon level (see, for example, Ref.~\cite{Bishara:2017pfq} and references therein).
In this work, we focus on scalar-mediated DM-quark and DM-gluon interactions, which could, for example, arise from the exchange of particles from
an extended Higgs sector in UV models.   However, we emphasize that our nuclear matrix elements apply also to the case of ``light'' scalar mediators,
with masses below the electroweak scale (the expression for the DM-nucleus scattering amplitude would have to be multiplied in that case by the
appropriate light scalar propagator).  The choice of scalar-mediated interactions  for this exploratory study is motivated by
the fact that two-nucleon currents arise in this case already at next-to-leading order (NLO) in the chiral counting, while they are relatively more suppressed for other interactions~\cite{Hoferichter:2015ipa}.

We focus on DM scattering off  a variety of light nuclei, namely $^2$H, $^3$H, $^3$He, $^4$He, and $^6$Li.
Our study has a twofold motivation. First,   for such light nuclei  first-principles calculations of the
nuclear wave functions are possible, once nucleon-level interactions are specified.
Therefore one can reliably study the effect of one- and two-nucleon currents for different spin and isospin structures.
Second,  light nuclear targets are of great interest because they provide a
better kinematic match for light DM and allow one to probe  sub-GeV  DM masses~\cite{Battaglieri:2017aum}.
In fact, both $^3$He and $^4$He isotopes are being considered for future direct detection
experiments~\cite{Guo:2013dt,Ito:2013cqa,Gerbier:2014jwa,Profumo:2015oya,Hertel2018}, including directional detection~\cite{Mayet:2016zxu}.
So our study goes beyond the benchmarking scope and will be relevant in the interpretation of results from these experiments.

In our study we follow a hybrid approach in which the scalar-mediated DM-nucleon  interactions  are derived in the framework of
chiral EFT up to NLO in the Weinberg counting~\cite{Cirigliano2012}, and the nuclear wave functions are obtained from
a phenomenological nuclear Hamiltonian that includes accurate two-body~\cite{Wiringa1995}
and three-body interactions~\cite{Pudliner1997}.
This allows us to take advantage of Quantum Monte Carlo methods, that in recent
years have proven to be extremely successful in describing light and
medium-heavy nuclei from first principles~\cite{Carlson:2015,Lynn:2016,Lonardoni:2018}.
Within  this framework,  the impact of  two-body currents has been previously
studied  in electron scattering~\cite{Lovato:2013,Lovato:2016}
and  neutral-current neutrino scattering~\cite{Lovato:2014,Lovato:2016}
[finding effects up to O(10\%)], 
 as well as in  $\beta$ decays~\cite{Pastore:2018}
(finding  effects of a few percent).

First-principles studies of DM-nucleus scattering for light nuclei already exist in the recent literature~\cite{Gazda:2016mrp,Korber2017}.
Reference~\cite{Korber2017} focuses on systems with $A=2$, and 3 and performs a self-consistent analysis of scalar-mediated
DM-nucleus scattering using both chiral currents and chiral potentials for the nuclear wave functions.
Reference~\cite{Gazda:2016mrp}, on the other hand,
focuses on $^3$He and $^4$He isotopes and
uses a hybrid approach (different from ours) in which
nuclear wave functions are obtained in the no-core shell model with next-to-next-to-leading-order chiral potential
while general  one-body ``currents'' (not just scalar-mediated) are parametrized in the non-relativistic EFT framework of Ref.~\cite{Fitzpatrick:2012ix}.
While overlapping with these studies, our work provides the first results for  two-nucleon currents in
systems with $A=4$, and 6, including the $^4$He isotope of experimental interest.

The paper is organized as follows:  we summarize the relevant scalar-mediated
DM-nucleon interactions in Sec.~\ref{sec:Scalar}.  In Sec.~\ref{sec:VMC} we give the details of the nuclear
Hamiltonian and wave functions used for the calculations of the elastic
scattering cross section and in Sec.~\ref{sec:Results} we present our results.
We give our conclusions and outlook in Sec.~\ref{sec:Conclusions}.

\section{\label{sec:Scalar}Scalar interaction}
A general, model-independent interaction for DM and quarks can be built
using higher dimension operators of the form (see for example Ref.~\cite{Bishara:2017pfq})
\begin{equation}
\mathcal{O} = \bar\chi\Gamma_\chi\chi \bar\psi\Gamma_\psi\psi,
\end{equation}
where $\Gamma_{\chi/\psi} \in \{\mathbbm{1}, \gamma^5,
\gamma^\mu, \gamma^\mu\gamma^5\}$ are Dirac bilinears,
$\chi$ and $\bar\chi$ are the DM fields, and $\psi$ and $\bar\psi$
the quark fields.

In this work, we restrict ourselves to scalar interaction between a
DM particle and standard model fields (vector and axial-vector interactions will
be studied in future work).  The DM particle is assumed to be a Dirac
fermion of spin-1/2.
The effective Lagrangian describing scalar-mediated DM-quark and DM-gluon  interactions is built from dimension-7
operators~\cite{Cirigliano2012}:
\begin{equation}
\mathcal L_{\text{eff}} =
\frac{1}{\tilde\Lambda^3} \left( \sum_{q=u,d,s} c_q \bar\chi\chi m_q \bar qq + c_G
\bar\chi\chi \alpha_s  G^a_{\mu\nu} G^{\mu\nu}_a \right),
\label{eq:lagrangian}
\end{equation}
where the sum runs over the light quark field $q$,   $\alpha_s$ is the strong coupling constant, and $G_{\mu \nu}$ is the gluon field strength tensor.
We have introduced a new physics scale, $\tilde\Lambda$, related to the
mass of the mediator (or possibly a new interaction mechanism) and
dimensionless Wilson coefficients $c_q$ and $c_G$ that parametrize the interaction.
For convenience, we include the masses of the quarks, $m_q$, in the
definition of the operators.

The derivation of the interaction at the nucleon level can be found
in Refs.~\cite{Cirigliano2012, Hoferichter2016, Bishara2017, Korber2017}.
The diagrams contributing at this order are shown in~\cref{fig:diagrams}.
Here we only summarize the resulting currents up to NLO\@,
in the context of $SU(2)$ chiral perturbation theory~\cite{Hoferichter2016,Korber2017}.

\newcommand{\figsize}{0.3}
\newcommand{\figscale}{0.6}

\begin{figure}[ht]
\centering
\begin{subfigure}{\figsize\linewidth}
\includegraphics[width=\columnwidth]{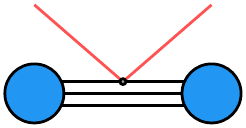}
\caption{}
\label{fig:lo}
\end{subfigure}
\begin{subfigure}{\figsize\linewidth}
\includegraphics[width=\columnwidth]{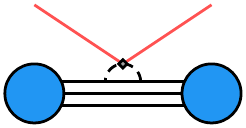}
\caption{}
\label{fig:nlo1}
\end{subfigure}
\begin{subfigure}{\figsize\linewidth}
\includegraphics[width=\columnwidth]{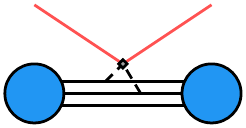}
\caption{}
\label{fig:nlo2}
\end{subfigure}
\caption{Diagrams contributing to DM-nucleus scattering
up to NLO\@. Solid black lines denote nucleons; dashed lines denote pions.
(a) Interaction at leading order (LO). (b) One-body interaction at NLO\@.
(c) Two-body interaction at NLO\@.
}
\label{fig:diagrams}
\end{figure}

We assume the following convention for momenta,
\begin{equation}
N(\vec p_i)+\chi(\vec k) \rightarrow N(\vec p_i')+\chi(\vec k'),
\end{equation}
where $\vec q = \vec k'-\vec k = \vec p_i-\vec p_i'$
and  $\vec p_i$ and  $\vec k$ ($\vec p_i'$ and $\vec k'$) are incoming
(outgoing) momenta for nucleons and DM particles, respectively
(the index $i$ refers to the $i$th nucleon).

In momentum space, the one body current describing the DM interaction
with the $i$th nucleon up to NLO  can be written as~\cite{Crivellin:2013ipa,Korber2017}

\begin{align}
\label{eq:curr_one}
J^{(1)}(\vec q_i) &= \frac{c_\text{is}}{\tilde \Lambda^3} \left[ \sigma_{\pi N}-\frac{9g_A^2\pi m_\pi^3}{4{(4\pi f_\pi)}^2} F\left( \frac{\lvert \vec q_i \rvert}{2m_\pi} \right) \right]
\nonumber \\ 
&- \frac{c_\text{iv}}{\tilde \Lambda^3} \frac{\delta m_N}{4} \tau_i^z
+ \frac{c_s}{\tilde \Lambda^3}  \left( \sigma_s  - \dot{\sigma}_s \vec{q}^2 \right)  - \frac{c_G}{\tilde \Lambda^3}  \ \frac{8 \pi   m_N^G}{9}
\nonumber \\
F(x) &= \frac{-x+(1+2x^2)\arctan{x}}{3x} \,,
\end{align}
where $\sigma_{\pi N}$ is the nucleon $\sigma$ term,   ${\delta m_N = {(m_n - m_p)}_{\rm strong}}$, ${\sigma_s = m_s  \langle N | \bar s s | N \rangle}$,   ${m_N^G = m_N - \sigma_{\pi N} - \sigma_s}$,
and ${\dot \sigma_s = (0.3 \pm 0.2) {\rm GeV}^{-2}}$~\cite{Hoferichter:2012wf}.
Moreover, we define the isoscalar and isovector couplings $c_{\rm is,iv}$  as the appropriate  linear combinations of the Wilson coefficients appearing in~\cref{eq:lagrangian}
\begin{align}
c_\text{is} &=\frac{c_u m_u+c_d m_d}{m_u+m_d},
 \\
c_\text{iv} &= 2\frac{c_d m_d-c_u m_u}{m_d-m_u}.
\end{align}
The numerical values for the single-nucleon quantities used in calculations are taken
from Refs.~\cite{Hoferichter2015PRL} and~\cite{Brantley2016}; i.e.,
\begin{equation}
\sigma_{\pi N} = (59.1\pm3.5)\ \text{MeV},\quad \delta m_N = (2.32\pm0.17)\ \text{MeV} \,.\\
\end{equation}

Even though we use the value for the $\sigma$ term obtained from a Roy-Steiner analysis of pion-nucleon scattering in Ref.~\cite{Hoferichter2015PRL}, our numerical results can be easily extended to other values coming, for example, from lattice QCD calculations (see Ref.~\cite{Shanahan:2016pla} and references therein).

As noted in Sec.~\ref{sec:Results}, the $\sigma$ term is factored out of the
cross section so the numerical input used will only affect the relative size
of the momentum-dependent part of the one-body current.

The two-body current appearing at NLO (\cref{fig:nlo2}), is given by
\begin{equation}
J^{(2)}_{\pi\pi}(\vec q_i, \vec q_j) =
-\frac{c_\text{is}}{\tilde\Lambda^3} {\left(\frac{g_A}{2F_\pi}\right)}^2 m_\pi^2
\vec{\tau}_i\cdot\vec{\tau}_j \frac{\vec{\sigma}_i\cdot\vec{q}_i
\vec{\sigma}_j\cdot\vec{q}_j}{(\vec{q}_i^2+m_\pi^2)(\vec{q}_j^2+m_\pi^2)} \,.
\label{eq:curr_two}
\end{equation}
The coordinate-space expressions of the currents are provided in the Appendix.
Two-nucleon currents proportional to $c_G$ appear formally at next-to-next-to-next-to-leading-order~\cite{Cirigliano2012,Crivellin:2013ipa,Korber2017}.

The elastic scattering cross section is given by
\begin{align}
\frac{d\sigma}{d\vec q^2} &= \frac{1}{4\pi v_\chi^2} \frac{1}{2j+1}
\nonumber
\\
&\times \sum_{m_j,m_j'=-j}^j {\left\lvert \bra{\psi_{jm_j'}} J(\vec q) \ket{\psi_{jm_j}}\right\rvert}^2~,\qquad
\label{eq:cross_section}
\end{align}
where $v_\chi$ is the velocity of the DM particle
and we are adopting normalization of nonrelativistic states for the DM particle and the nucleus.
The nuclear matrix element for a given nucleus
with ground state $ \ket{\psi_{jm_j}}$ is characterized by total spin  $j$ and  spin polarization $m_j$ and is calculated using
$J(\vec q)$, given by the sum of one- and two-body contributions from Eqs.~\ref{eq:curr_one} and~\ref{eq:curr_two}.

\section{\label{sec:VMC}Nuclear Wave Functions}
The evaluation of nuclear matrix elements required in~\cref{eq:cross_section}
is performed using the variational Monte Carlo method.  We use
variational wave functions $\ket{\psi}$ that minimize the expectation
value of
\begin{equation}
E_V = \frac{\bra{\psi}H\ket{\psi}}{\braket{\psi}{\psi}} \,,
\end{equation}
which provides  an upper bound to the energy of the ground state.

The phenomenological Hamiltonian used in this work has an Argonne $v_{18}$
potential~\cite{Wiringa1995} for the two-body interaction and an Urbana
IX potential~\cite{Pudliner1997} for the three-body interaction:
\begin{equation}
H = \sum_i T_i + \sum_{i<j} v_{ij} + \sum_{i<j<k} v_{ijk} \,.
\end{equation}

The variational wave function for a given nucleus in the $J$ state is:
\begin{equation}
\ket{\psi} = \left[\mathcal{S}\prod_{i<j}^A (1+U_{ij})\right]
\left[\prod_{i<j<k} f_c(r_{ijk})\right] \ket{\Phi(JMTT_3)} \,,
\label{eq:wavefunction}
\end{equation}
where $\mathcal S$ is a symmetrization operator acting on two-
and three-body correlation operators, $f_c$ is a spin- and
isospin-independent two- and
three-body correlation, $\Phi$ is an antisymmetric wave function containing the
correct quantum numbers for the state of interest, and
the two-body spin- and isospin-dependent correlations are
constructed as
\begin{equation}
U_{ij} = \sum_p f^p(r_{ij}) O^p_{ij} \,,
\end{equation}
where the operators are
\begin{equation}
O^p_{ij} = \vec{\tau}_i\cdot\vec{\tau}_j,\vec{\sigma}_i\cdot\vec{\sigma}_j,
(\vec{\tau}_i\cdot\vec{\tau}_j)(\vec{\sigma}_i\cdot\vec{\sigma}_j),S_{ij},S_{ij}\vec{\tau}_i\cdot\vec{\tau}_j,
\end{equation}
and $f^p$ are radial functions. For more details see Ref.~\cite{Carlson:2015} and
references therein.

Finally, the currents entering~\cref{eq:cross_section} are given
by
\begin{equation}
J(\vec q) = \sum_i e^{i\vec q \cdot \vec r_i} J^{(1)}(\vec q) + \sum_{i<j} J^{(2)}_{\pi\pi}(\vec q; \vec r_i, \vec r_j)~,
\end{equation}
obtained by Fourier transforming the expressions in Eqs.~\ref{eq:curr_one} and~\ref{eq:curr_two},
as reported in the Appendix.

\section{\label{sec:Results}Results}
Here we present the results of our calculations, for a variety of light
nuclei.    Considering for the moment only the isoscalar part
[the contributions of $c_\text{iv}$, $c_s$ and $c_G$ are easily included according to~\cref{eq:rescaling} below],
it is convenient, as in Ref.~\cite{Korber2017}, to expand the total
cross section in terms of nuclear response functions:
\begin{equation}
\label{eq:structure}
\frac{d\sigma}{d\vec q^2} = \frac{c_\text{is}^2}{\tilde \Lambda^6} \,     \frac{\sigma^2_{\pi N} A^2}{4\pi v^2_\chi} {\left\lvert \mathcal{F}^{(0)}_\text{is}(\vec q^2) + \mathcal{F}^{(1)}_\text{is,2b}(\vec q^2) + \mathcal{F}^{(1)}_\text{is,r}(\vec q^2) \right\rvert}^2 \,,
\end{equation}
where we factorized the isoscalar coupling, $\sigma$ term, and the number of nucleons $A$.
Each function $\mathcal F^{(\nu)}_{a,i}$
carries the index $\nu$ referring
to the chiral order, the label $a$ to distinguish between isoscalar and isovector
contributions, and the label $i$ for contributions of two-body currents and for the so-called  ``nucleon radius'' correction, given by the one-body momentum-dependent correction in~\cref{eq:curr_one} proportional to $F\left( \frac{\lvert \vec q_i \rvert}{2m_\pi} \right)$.
With our choice of normalization, we have  $\mathcal{F}^{(0)}_\text{is}(0)=1$.

In what follows we concentrate  on the case $c_\text{is} \neq 0$ while setting  $c_\text{iv,s,G}/c_\text{is}=0$,
because to the order we work the additional couplings do not introduce independent nuclear responses.
In fact, from Eq.~\eqref{eq:curr_one} one can obtain the cross section for general couplings $c_\text{iv,s,G} \neq 0$
by rescaling  $\mathcal{F}^{(0)}_\text{is}(\vec{q}^2)$  in  Eq.~\eqref{eq:structure}   by the factor
\begin{align}
\label{eq:rescaling}
1 &- \left(\frac{c_\text{iv}}{c_\text{is}}\right) \frac{\delta m_N}{4\sigma_{\pi N}} \frac{2 Z - A}{A} + \left( \frac{c_\text{s}}{c_\text{is}} \right) \frac{\sigma_s  - \dot \sigma_s \, \vec{q}^2}{\sigma_{\pi N}}\\
\nonumber
&-\left( \frac{c_G}{c_\text{is}} \right) \frac{8 \pi m_N^G}{9 \sigma_{\pi N}}~.
\end{align}

\newcommand{\figsizeb}{0.4}

\begin{figure*}[ht]
\begin{subfigure}{\figsizeb\linewidth}
\includegraphics[width=\columnwidth]{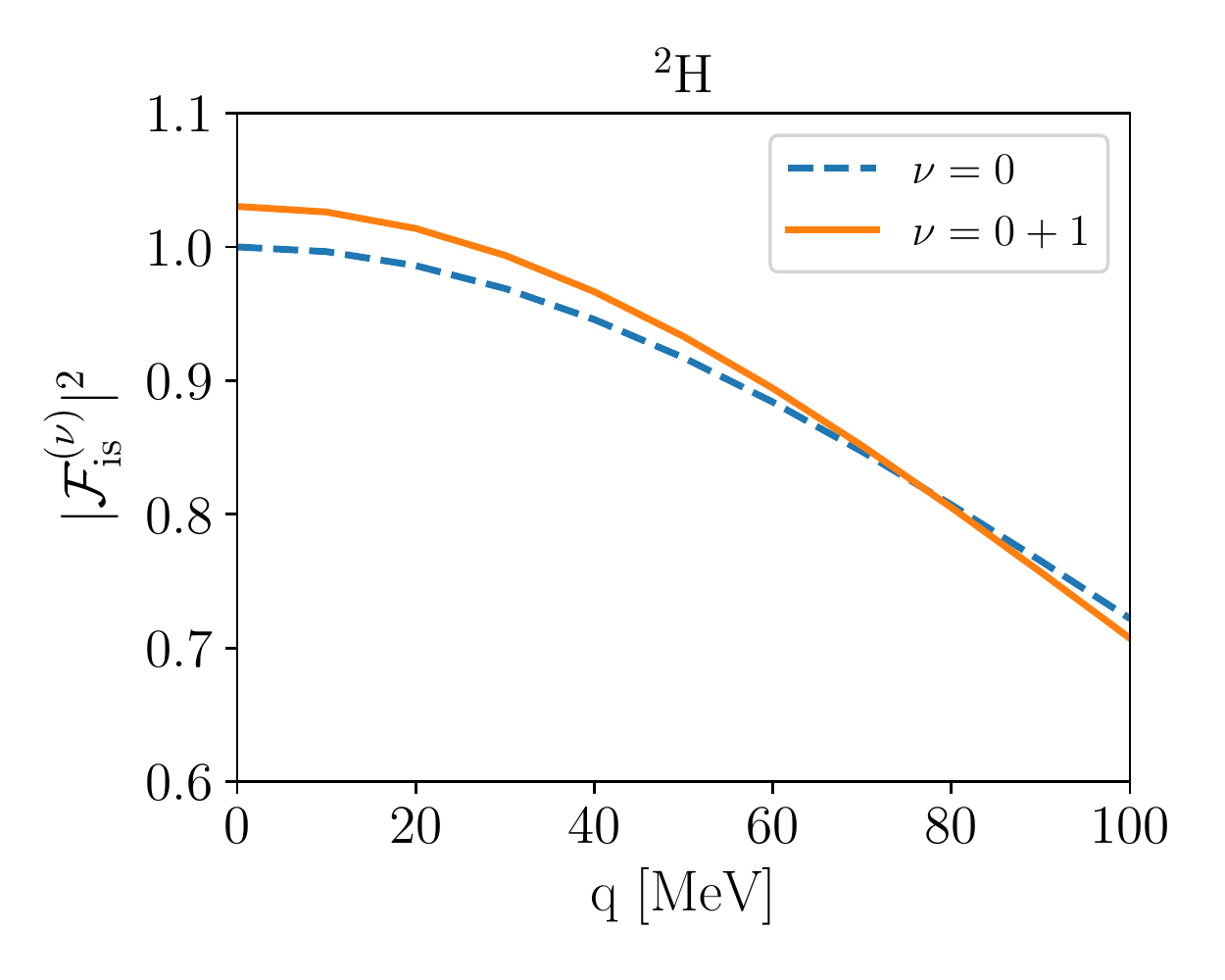}
\caption{}
\end{subfigure}
\begin{subfigure}{\figsizeb\linewidth}
\includegraphics[width=\columnwidth]{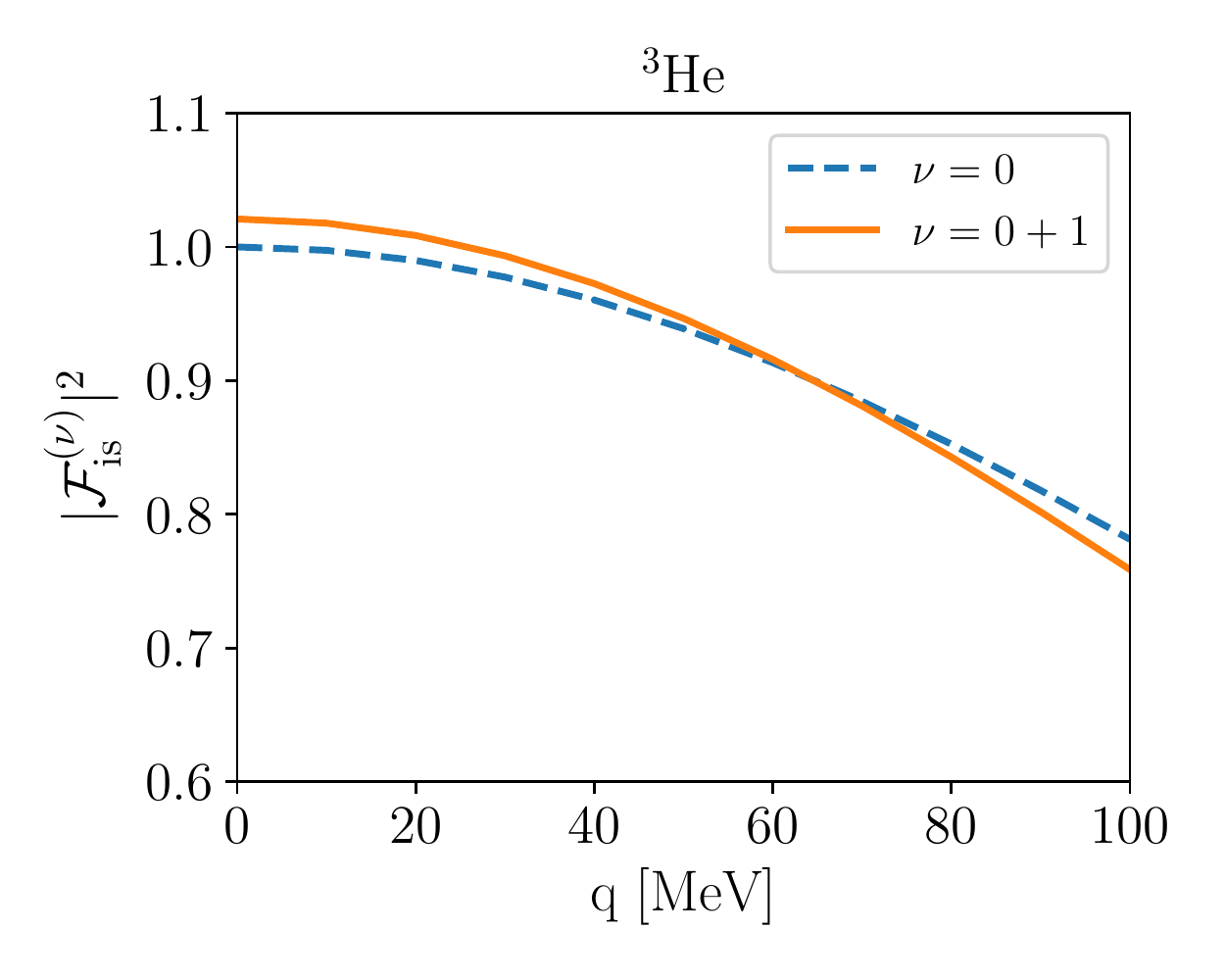}
\caption{}
\end{subfigure}
\\
\begin{subfigure}{\figsizeb\linewidth}
\includegraphics[width=\columnwidth]{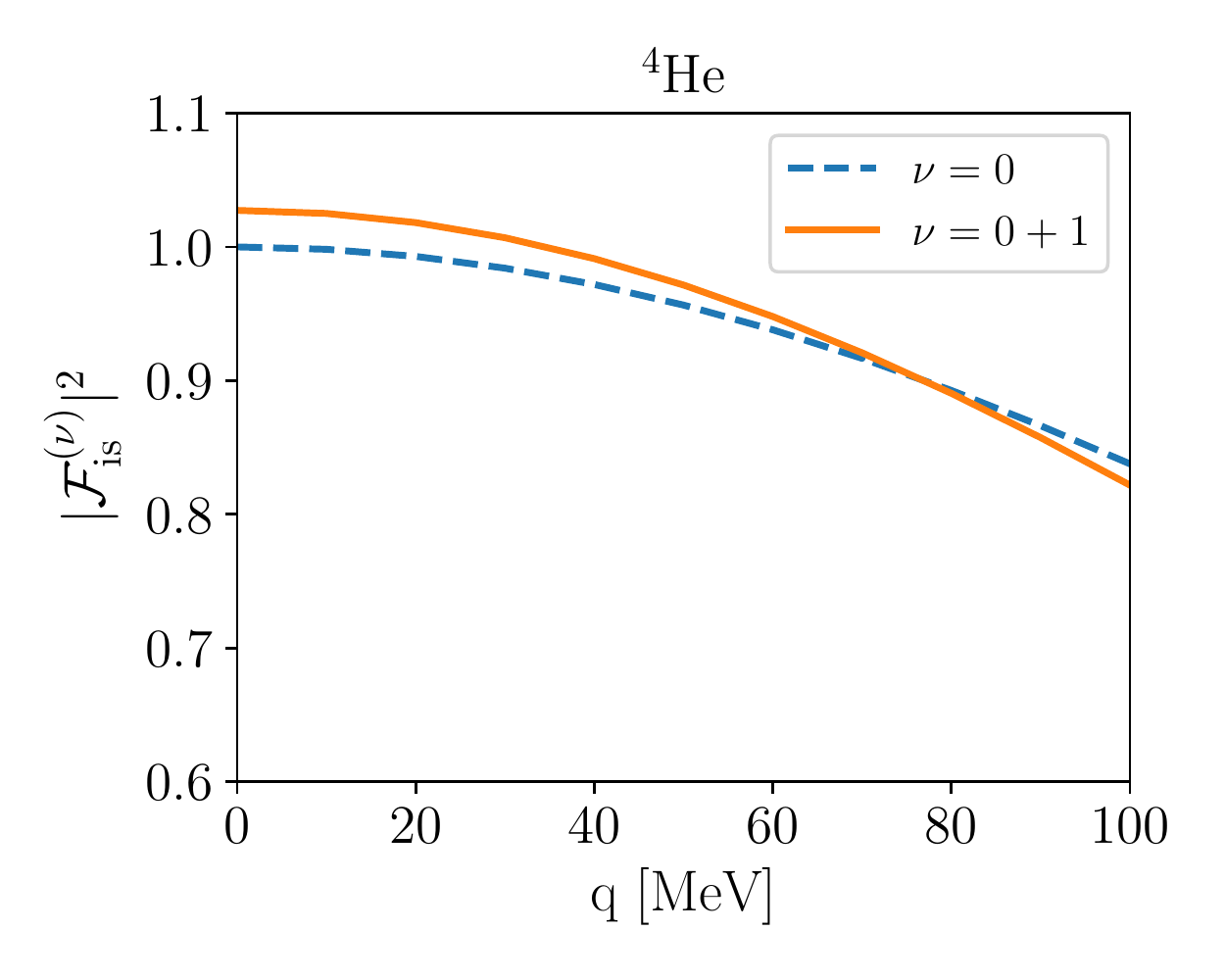}
\caption{}
\end{subfigure}
\begin{subfigure}{\figsizeb\linewidth}
\includegraphics[width=\columnwidth]{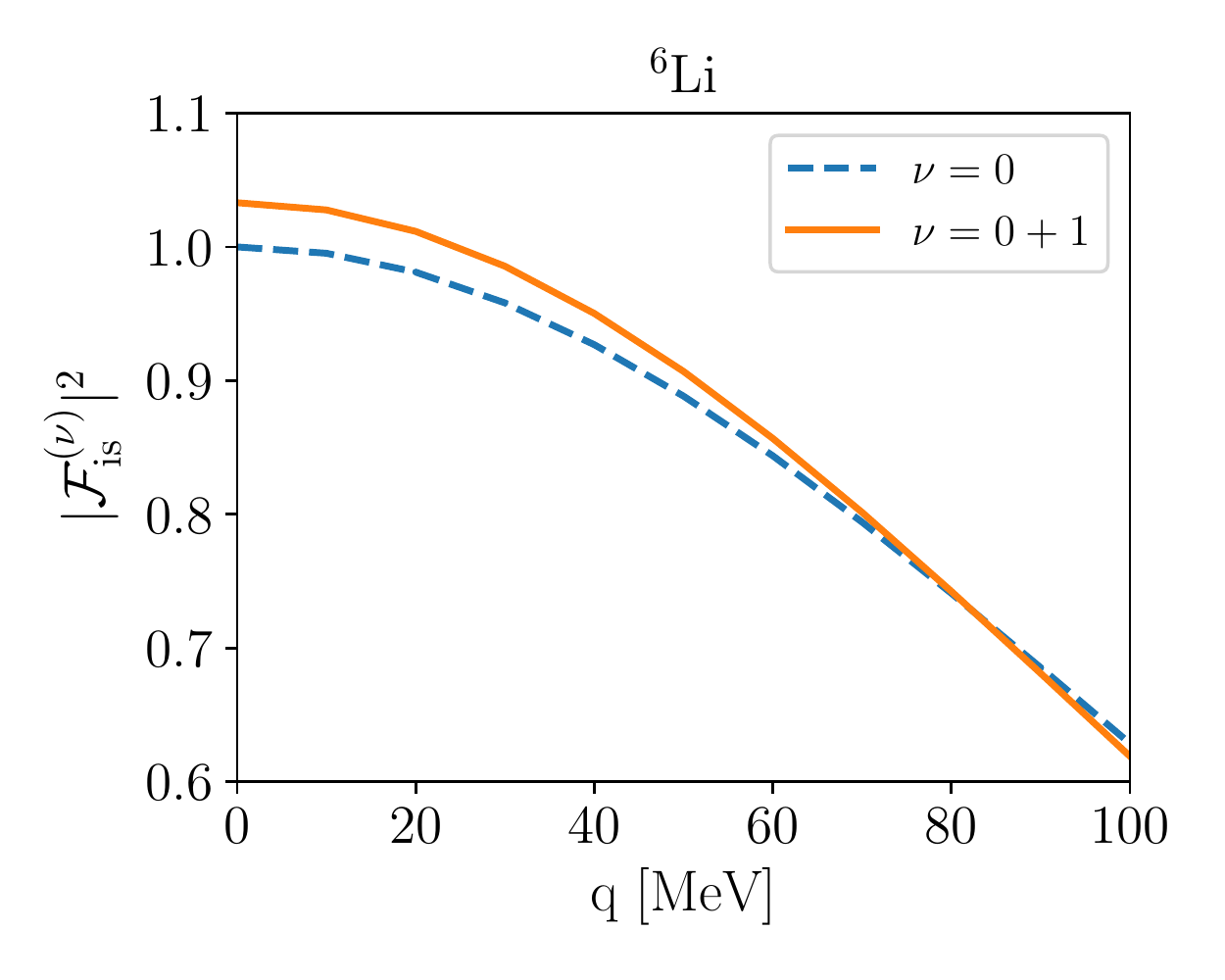}
\caption{}
\end{subfigure}
\caption{Isoscalar matrix elements for nuclei from $A=2$ to 6.
Dashed blue lines correspond to LO calculations, and orange solid ones correspond
to NLO.}
\label{fig:cross}
\end{figure*}

The maximum momentum transfer $q$ considered in the calculations
is 100 MeV, which is appropriate for light nuclei and a
DM mass of about 1 GeV. In this scenario $q$ ranges from a few to tens of MeV.
In~\cref{fig:cross} we present the results for  isoscalar terms in~\cref{eq:structure}.
For each nucleus, we compare the results for  LO and NLO\@ contributions.
As we can see, the order ($\nu=1$) corrections slightly increase the cross section at low momenta.
At larger momenta, the contribution from the radius correction is greater than the
two-body contribution and of opposite sign, making the total cross section decrease as $q$ increases.
This behavior is consistent for all the nuclei considered here.
Nonetheless, in the range of values considered, the deviation from LO results is at the few percent level.

\begin{figure}[ht]
\includegraphics[width=\columnwidth]{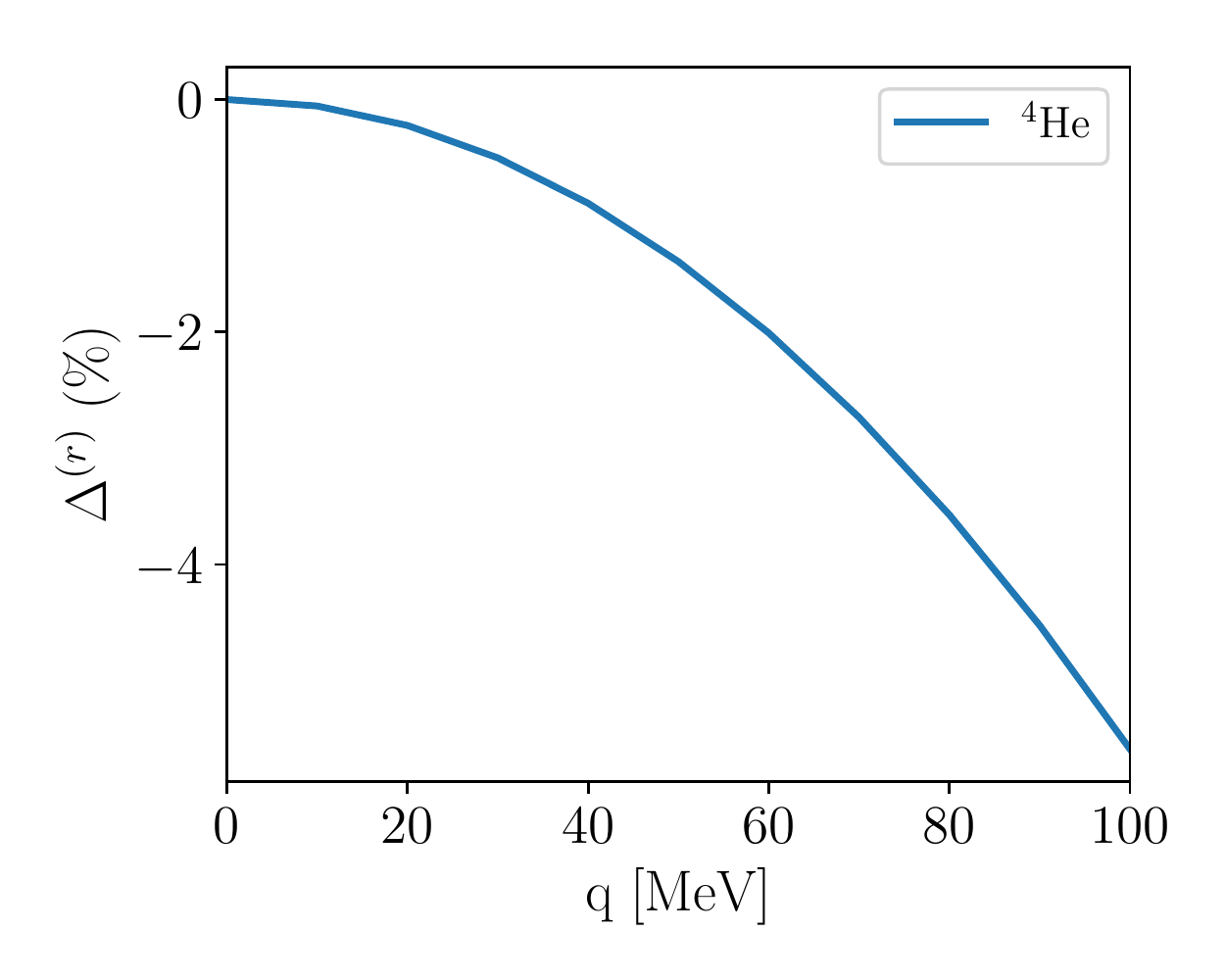}
\caption{Percentual radius correction for $^4$He}
\label{fig:dr}
\end{figure}

To assess the effect of   the two terms appearing at order ($\nu=1$), it is
useful to consider their relative contribution to the total cross section.
First,  we define the radius correction in the following way~\cite{Korber2017}
\begin{equation}
\Delta^{(r)} = \frac{{| \mathcal{F}^{(0+1)}_\text{is}(\vec q^2)|}^2
- {| \mathcal{F}^{(0)}_\text{is}(\vec q^2) + \mathcal{F}^{(1)}_\text{is,2b}(\vec q^2) |}^2}
{{| \mathcal{F}^{(0+1)}_\text{is}(\vec q^2)|}^2}~,
\label{eq:rad}
\end{equation}
where
$\mathcal{F}^{(0+1)}_\text{is}(\vec q^2)$ is defined by the sum of the three
isoscalar terms on the right-hand side of~\cref{eq:structure}.
Working at NLO and expanding for small order ($\nu=1$) corrections,
this expression reduces to
\begin{align}
\Delta^{(r)} \sim \frac{2 \mathcal{F}^{(1)}_\text{is,r}(\vec q^2)}
{\mathcal{F}^{(0+1)}_\text{is}(\vec q^2)} \sim -\frac{2}{\sigma_{\pi N}}
\frac{9g_A^2\pi m_\pi^3}{4{(4\pi f_\pi)}^2} F\left( \frac{\abs{\vec q}}{2m_\pi} \right).
\label{eq:radius_expansion}
\end{align}
The nuclear effects drop out and the correction is given only by the
momentum-dependence of~\cref{eq:curr_one}.  This expression agrees
with the complete nuclear calculations, in the range of the momenta
considered here.  For this reason, we only present the radius correction
for $^4$He in~\cref{fig:dr},
and  note that  all the other nuclei show the same behavior,
up to  minor  differences due to  the two-body contribution and higher order terms in the expansion
(\ref{eq:radius_expansion}).
The radius correction vanishes at zero momentum transfer and grows to
about 6\%  at $q=100\ \text{MeV}$.

\begin{figure}[htb]
\includegraphics[width=\columnwidth]{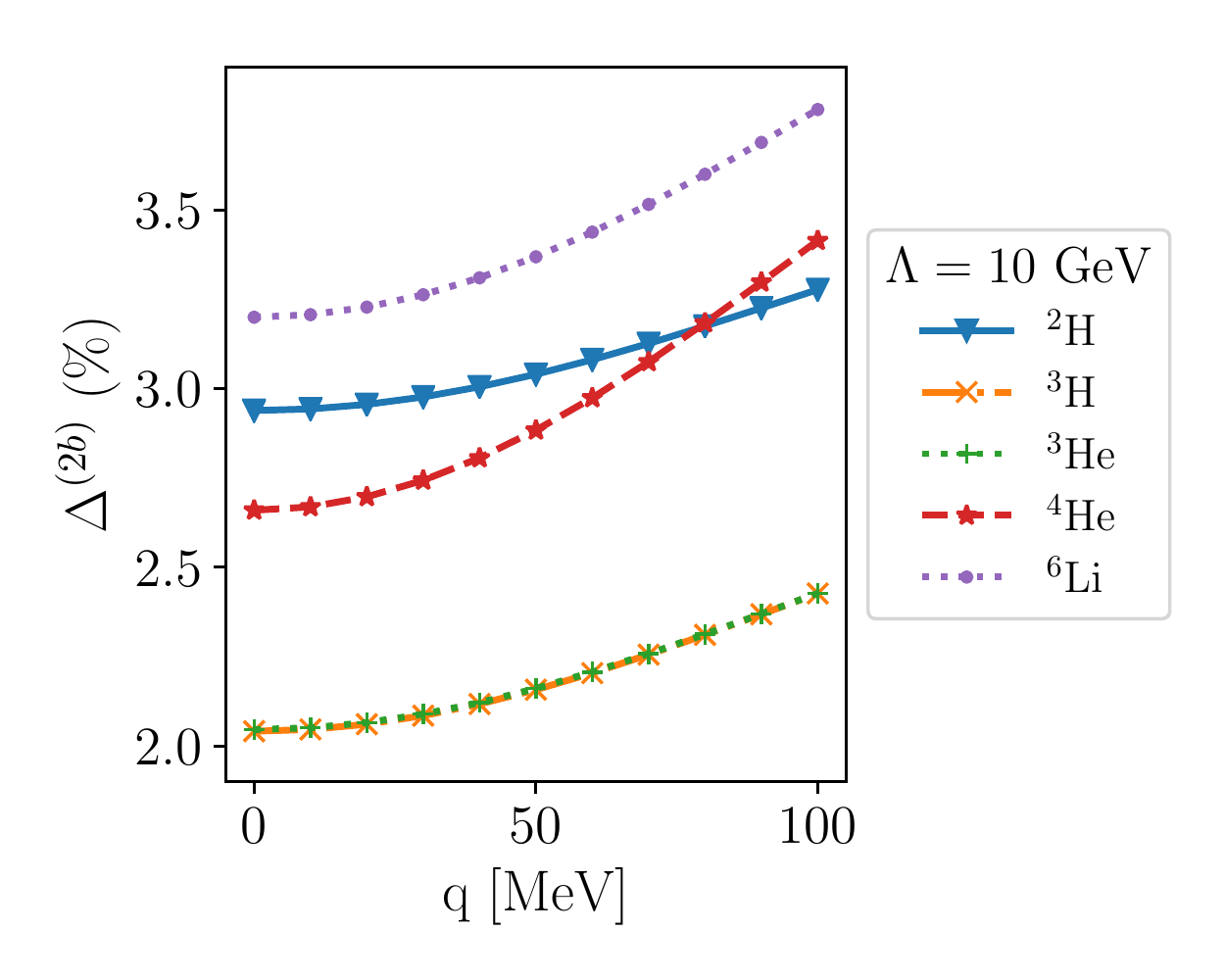}
\caption{Percentual two-body correction to the total cross section
for various nuclei.}
\label{fig:delta2b}
\end{figure}

Similarly to~\cref{eq:rad}, the relative contribution of two body
currents is given by~\cite{Korber2017}
\begin{align}
\Delta^{(2b)} &= \frac{{| \mathcal{F}^{(0+1)}_\text{is}(\vec q^2)|}^2
- {| \mathcal{F}^{(0)}_\text{is}(\vec q^2) + \mathcal{F}^{(1)}_\text{is,r}(\vec q^2) |}^2}
{{| \mathcal{F}^{(0+1)}_\text{is}(\vec q^2)|}^2}.
\label{eq:twob}
\end{align}
In~\cref{fig:delta2b} we present the percentual correction
given by two-body operators entering at NLO with respect to the
total contribution up to NLO\@.
All two-body corrections are of modest size, with nuclei with $A=3$
giving a smaller contribution compared to $^2$H and $^4$He, and being
almost exactly equal.
The two-body corrections tend to increase
with the nucleus size at large momenta
and this effect might be even more pronounced for larger nuclei.
We notice however
that at low momenta the correction in the $^2$H nucleus is somehow larger
than $A=3$ and 4 nuclei.
Overall, the role of two-body operators increases with the momentum transferred,
from about 2$\%$ up to about 4$\%$.
This is true only for very large cutoff. Note, however,
that the actual size of the correction depends on our choice
$c_\text{iv,s,G}=0$,
and can be computed in the general case  through the rescaling  introduced
in Eq.~\eqref{eq:rescaling} above.
Also, radius and two-body corrections for different values of the nucleon $\sigma$ term can be obtained from our data by multiplying by the appropriate constant. Lower values of the $\sigma$ term as in Ref.~\cite{Shanahan:2016pla} increase the relative size of NLO contributions.

\begin{figure}[htb]
\includegraphics[width=\columnwidth]{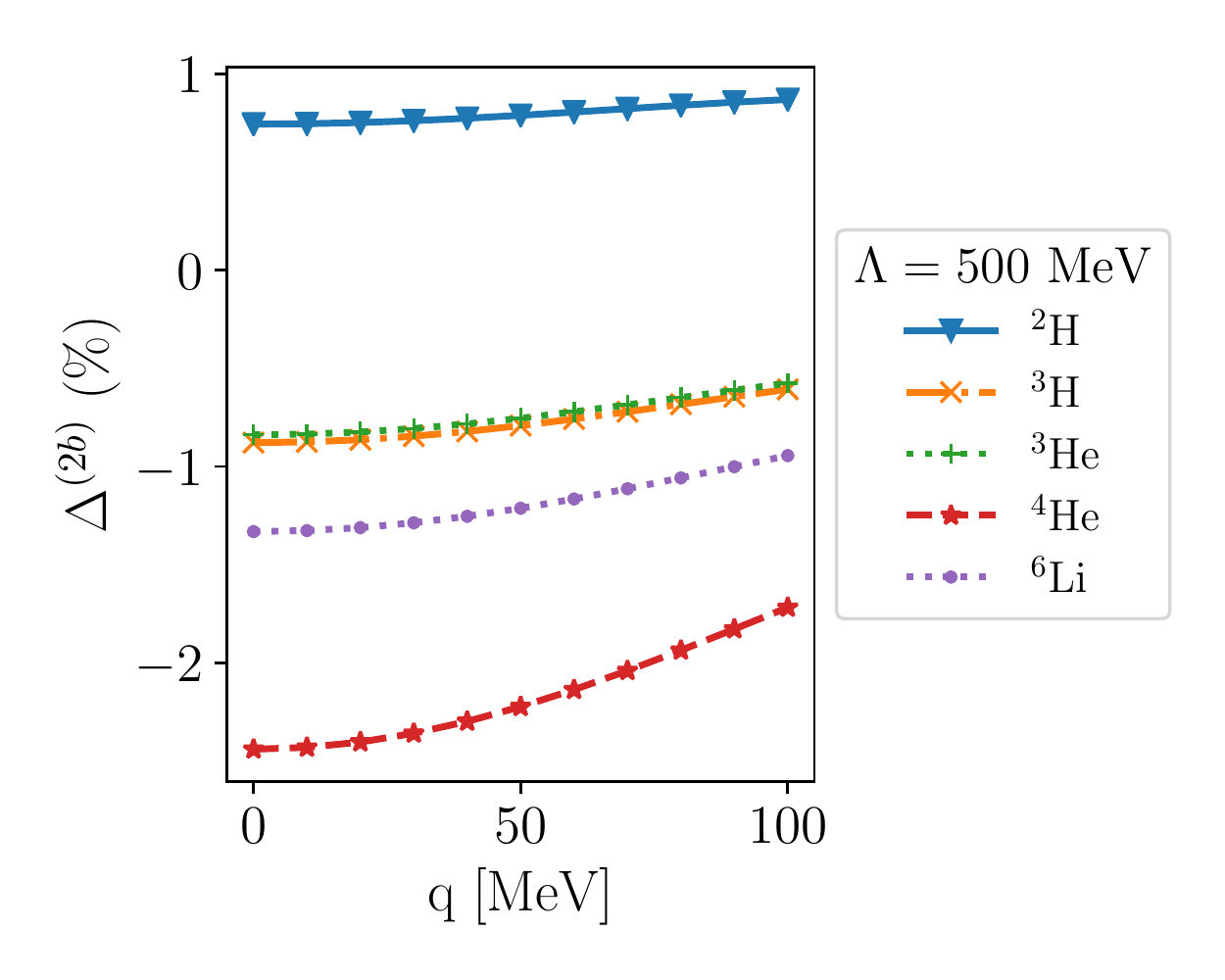}
\caption{(color online) Percentual two-body correction to the total cross section
for various nuclei.}
\label{fig:delta2b_500}
\end{figure}

\begin{figure}[t]
\includegraphics[width=\columnwidth]{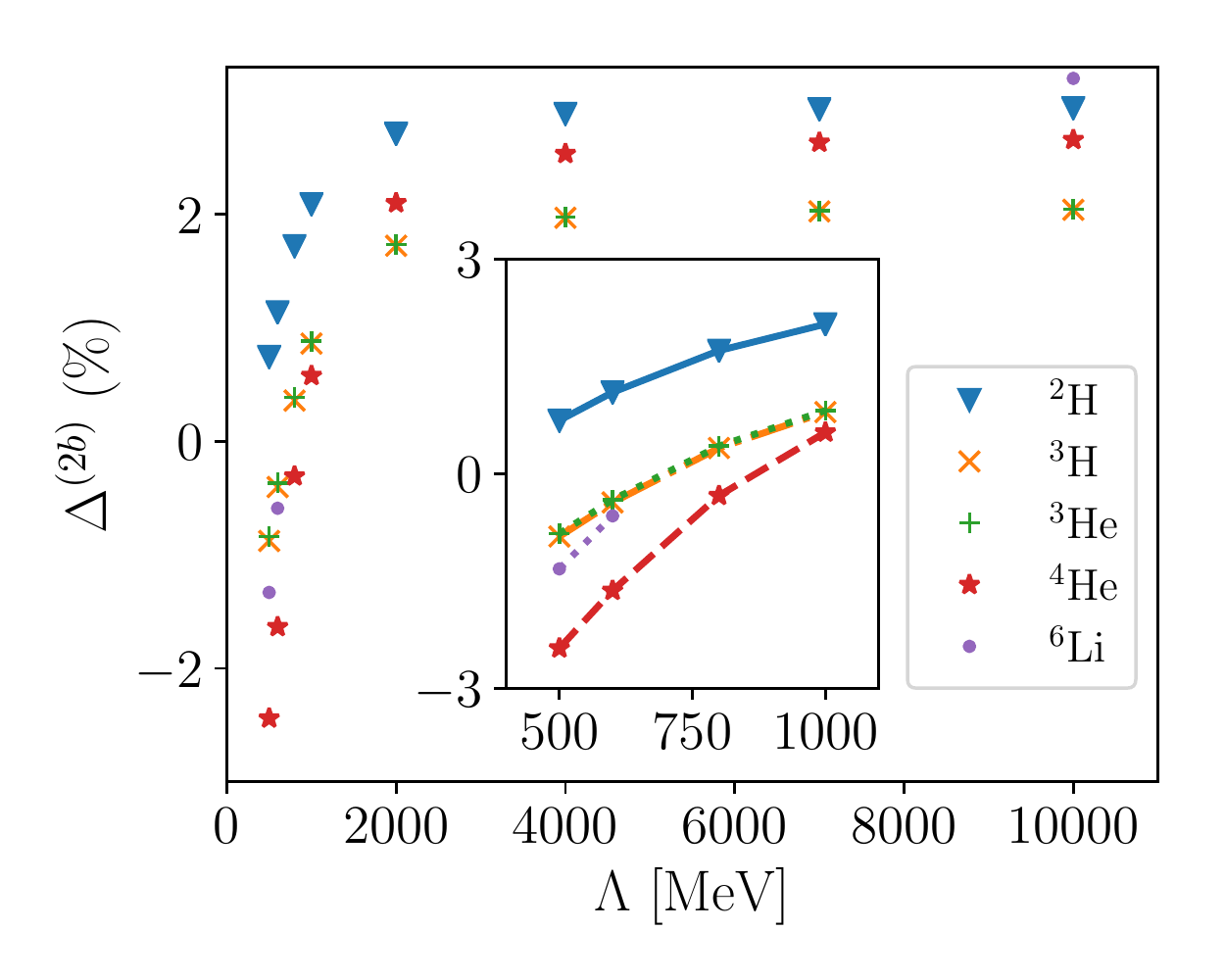}
\caption{Cutoff dependence of the two-body contribution for $q=0$.}
\label{fig:2b_lambda}
\end{figure}

Finally, we discuss the  cutoff-dependence of the nuclear matrix elements
due to the short-distance regulator introduced in the Fourier transforms (see the Appendix).
All the results reported so far were obtained in the limit of infinite cutoff
$\Lambda$ in~\cref{eq:twob_coord}.
Ideally, one should consider a cutoff in the current consistent  to
the one used in the nuclear Hamiltonian, but this is not possible in our
hybrid approach. In fact,   because we use a phenomenological potential in the nuclear
Hamiltonian, there are no ``strong''  low-energy constants that allow for a variation of the
cutoff when obtaining the nuclear wave function.
In practice,   because the Argonne $v_{18}$ interaction has a very strong hard core,
we might expect its effective cutoff to be very high.
In such situations,   a possible strategy
would be to fix the cutoff in the currents, fit the ``weak''  low-energy constants to reproduce some
observable, and predict properties of larger nuclei.
For Argonne Hamiltonians this has been for example explored
in $\beta$-decay calculations~\cite{Pastore:2018}.
However, in the present case, up to the order we work, there are no new
low-energy constants in the currents and this approach is not viable.
So  to explore the cutoff dependence we have simply calculated $\Delta^{(2b)}$
for different values of $\Lambda$.
The calculations are presented in Figs.~\ref{fig:delta2b} and~\ref{fig:delta2b_500} where we show the fractional two-body corrections at $\Lambda = 500$~MeV and $\Lambda = 10$~GeV, respectively, as a function of $q$. In~\cref{fig:2b_lambda} we show the two-body corrections for all the nuclei considered here as a function of the cutoff $\Lambda$ for a fixed $q=0$.

\begin{table}[htb]
\begin{center}
\begin{ruledtabular}
\begin{tabular}{ccccc}
Nucleus & $\Lambda$ (MeV) & & $\Delta^{(2b)}\ (\%)$ &\\
& & $\mathcal O_1$ & $\mathcal O_2$ & $\mathcal O_1 + \mathcal O_2$\\
\colrule 
$^2$H & 500 & -3.4 & 4.2 & 0.7\\
      & 10000 & -4.9 & 7.9 & 3.0\\
$^4$He & 500 & -13.2 & 10.7 & -2.4\\
       & 10000 & -19.2 & 21.9 & 2.7\\
\end{tabular}
\end{ruledtabular}
\caption{Percentual two-body correction to the total cross section for nonvanishing operators contributing at $q=0$.}
\label{tab:cancellation}
\end{center}
\end{table}

Two features emerge from our results.
First,  for  $\Lambda \in [500, 1000]$~MeV  there is a strong cutoff dependence
of the two-body contribution, so that it even changes signs for some nuclei.
For example, the two-body correction
is always positive for $^2$H, but it changes sign in $A=3$
nuclei and $^6$Li for $\Lambda\sim700$~MeV and above 800 MeV for
$^4$He.
This is due to the fact that there is a large cancellation between the operators in~\cref{eq:twob_coord}.
We illustrate this point by reporting in~\cref{tab:cancellation} the fractional contributions to the total cross section at $q=0$, which arise entirely from the operators
$\mathcal O_1$ ($\sim \vec\sigma_1\cdot\vec\sigma_2$) and
$\mathcal O_2$ ($\sim \vec \sigma_1\cdot \hat{\vec r}\; \vec \sigma_2\cdot \hat{\vec r}$) in~\cref{eq:twob_coord}.
The second feature is that the two-body contribution saturates
for large values of $\Lambda$, starting around 2 GeV. This might
reflect the fact that the phenomenological nuclear Hamiltonian
considered here effectively has a very large cutoff.
Overall, the cutoff dependence of the two-body current contribution
is the largest source of   uncertainty in our approach.
Attempts to remove this ``systematic'' effect will necessarily involve
the use of wave functions obtained by a chiral potential, as discussed in Ref.~\cite{Korber2017}.

\section{\label{sec:Conclusions}Conclusions}

We have studied the  elastic scattering of DM particles off a number of   light nuclei
($^2$H, $^3$H, $^3$He, $^4$He, and $^6$Li)  with different spin and isospin
using quantum Monte Carlo methods.  We have focused on scalar-mediated interactions,
parametrized by four Wilson coefficients  related to the mediator mass  and its coupling to DM and quarks.
We have used the resulting hadronic currents up to NLO in the chiral expansion,
containing both  nucleon ``scalar radius'' corrections and two-body effects.
We have followed a hybrid approach in which   the chiral EFT currents  are used in combination with
 nuclear wave functions  obtained from a phenomenological nuclear Hamiltonian
 that includes the Argonne $v_{18}$ two-body interaction, and the three-body Urbana IX interaction.

We find that for the momentum transfers of interest,   the overall size of the NLO  corrections is at the few percent level,
perhaps smaller than suggested by chiral counting.
The  NLO correction due to the nucleon scalar radius  is essentially free of nuclear structure uncertainties
and grows from zero to $\approx- 2\%$ at $q = 60$~MeV  and  $\approx -6\%$ at $q = 100$~MeV.
On the other hand, the corrections due to scalar two-body currents
 --- estimated for $A=4$ and 6 for the first time in this work ---
start at $q=0$ at  the 2--3\% level  (depending on the nucleus)
and mildly grow with $q$.
For $A=2$ and 3, our results are in qualitative agreement with~\cite{Korber2017}.

We can also compare our findings for $^3$He and $^4$He with those of Ref.~\cite{Gazda:2016mrp}.
This reference considers only  the one-body current,  generated by the operator $\hat{\mathcal O}_1$ in the NREFT
operator basis of Ref.~\cite{Fitzpatrick:2012ix}.
While a detailed  numerical comparison  is  beyond the scope of our work, for the one-body contribution
we find a good qualitative agreement with the results of Ref.~\cite{Gazda:2016mrp}.

Even assigning a conservative uncertainty as large as the
variation of the two-body matrix element between $\Lambda = 500$~MeV and
$\Lambda = 2$~GeV,
the total cross-section is still known
quite precisely, namely at the few percent level.
Therefore,  our results  in combination with Refs.~\cite{Gazda:2016mrp,Korber2017}
already  provide the reasonable  nuclear structure input  needed to assess the  sensitivity of future
experimental searches of  light dark matter  using $^3$He and $^4$He targets.

Further refinements are certainly warranted.   Interesting directions for future studies
include:  (i) moving beyond the hybrid approach, in the spirit of Ref.~\cite{Korber2017},
by using chiral interactions (as opposed to the  Argonne $v_{18}$ potential)
 in combination with Quantum Monte Carlo to obtain the nuclear wave functions;
and (ii) exploring the consistency of Weinberg power counting in various channels
of   DM-nucleon  two-body interactions, and matching to lattice QCD calculations~\cite{PhysRevLett.120.152002,Beane:2013kca}, to determine the relevant low-energy constants

\begin{center}
{\bf  Acknowledgments}
\end{center}
We thank J. Carlson, W. Detmold, D. Lonardoni, E. Mereghetti, S. Pastore, R. Schiavilla and J. de Vries for  discussions and correspondence.
V.C.  acknowledges support by the U.S. DOE Office of Nuclear Physics and by the LDRD program at Los Alamos National Laboratory.
The work of S.G. was supported by the NUCLEI
SciDAC program, by the U.S. DOE under contract
DE-AC52-06NA25396, 
by the LANL LDRD program, and by
the DOE Early Career Research Program.
Computational resources
have also been provided by Los Alamos Institutional Computing,
and we also used resources provided by the
NERSC, which is supported by the U.S. DOE under Contract No. DE-AC02-05CH11231.

\onecolumngrid 
\begin{appendices}
\setcounter{equation}{0}
\renewcommand{\theequation}{A\arabic{equation}}
\section*{\label{sec:Appendix}Appendix}
The one- and two-body DM-nucleon currents need  to be Fourier transformed so that they can be used in a variational Monte Carlo calculation in coordinate space.
To tame the short-distance singularities we use  a Gaussian regulator of the form
\begin{equation}
S_\Lambda(\vec k^2) = e^{-\frac{\vec k^2}{2\Lambda^2}}~,
\end{equation}
with cutoff parameter $\Lambda$.
The two-body current is obtained from
\begin{equation}
J^{(2)}_{\pi\pi}(\vec q; \vec r_1, \vec r_2) =  \int \frac{d^3 \vec k_1}{{(2\pi)}^3} \frac{d^3 \vec k_2}{{(2\pi)}^3} e^{i\vec{k}_1\cdot\vec{r}_1} e^{i\vec{k}_2\cdot\vec{r}_2} S_\Lambda(\vec k_1^2) S_\Lambda(\vec k_2^2)
{(2\pi)}^3 \delta^{(3)}(\vec k_1 + \vec k_2 - \vec q) J^{(2)}_{\pi\pi}(\vec k_1, \vec k_2)
\end{equation}
The coordinate space expression for two-body currents can be calculated analytically, except for one integration over an auxiliary variable $y$.
It reads
\begin{align}
\begin{aligned}
	J^{(2)}_{\pi\pi}(\vec q;\vec r_1, \vec r_2) = &-\frac{1}{\Lambda^3} {\left(\frac{g_A}{2F_\pi}\right)}^2 c_\text{is}m_\pi^2 \vec{\tau}_1\cdot\vec{\tau}_2
	\frac{1}{2} e^{i\vec{q}\cdot\vec{R}} \int_{-1}^{1}dy e^{-i\vec{q}\cdot\vec{r}y/2}\\
	&\times \Bigg[
	(\vec{\sigma}_1\cdot\vec{q})(\vec{\sigma}_2\cdot\vec{q}) \frac{1-y^2}{4} s(r,y) +
	(\vec{\sigma}_1\cdot\vec{q})(\vec{\sigma}_2\cdot\hat{\vec{r}}) \left( -i \frac{1+y}{2} \right) \frac{\partial}{\partial r} s(r,y)\\
	&+(\vec{\sigma}_1\cdot\hat{\vec r})(\vec{\sigma}_2\cdot\vec{q}) \left( i \frac{1-y}{2} \right) \frac{\partial}{\partial r} s(r,y) +
	(\vec{\sigma}_1\cdot\vec{\sigma}_2) \frac{1}{r} \frac{\partial}{\partial r} s(r,y) +
	(\vec{\sigma}_1\cdot\hat{\vec r})(\vec{\sigma}_2\cdot\hat{\vec r}) r\frac{\partial}{\partial r} \frac{1}{r} \frac{\partial}{\partial r} s(r,y) \Bigg],
	\label{eq:twob_coord}
\end{aligned}
\end{align}
where $\vec r = \vec r_2 - \vec r_1$, $\vec R = \frac{\vec r_2 + \vec r_1}{2}$, and the radial functions have the following expressions:

\begin{align}
s(r,y) = &\frac{e^{L^2/\Lambda^2}}{8\pi L \Lambda r}\Bigg[ \text{erfc}\left( \frac{L}{\Lambda} + \frac{\Lambda r}{2} \right) e^{Lr} \left( \frac{L}{\Lambda} + \frac{\Lambda r}{2} \right) - \text{erfc}\left( \frac{L}{\Lambda}-\frac{\Lambda r}{2} \right) e^{-Lr} \left( \frac{L}{\Lambda}-\frac{\Lambda r}{2} \right) \Bigg],\\
    \nonumber
\frac{\partial}{\partial r} s(r,y) = &\frac{e^{L^2/\Lambda^2}}{8\pi \Lambda^2 r^2}\Bigg[ \text{erfc}\left( \frac{L}{\Lambda} + \frac{\Lambda r}{2} \right) e^{Lr} \left( - 1 + Lr + \frac{\Lambda^2 r^2}{2} \right)\\
     &+\text{erfc}\left( \frac{L}{\Lambda}-\frac{\Lambda r}{2} \right) e^{-Lr} \left( 1 + Lr - \frac{\Lambda^2 r^2}{2} \right) \Bigg] - \frac{e^{-\Lambda^2 r^2/4}}{4\pi^{3/2}\Lambda r},\\
    \nonumber
	r\frac{\partial}{\partial r} \frac{1}{r} \frac{\partial}{\partial r} s(r,y) = &\frac{e^{L^2/\Lambda^2}}{8\pi \Lambda^2 r^3} \Bigg[ \text{erfc} \left( \frac{L}{\Lambda} + \frac{\Lambda r}{2} \right) e^{Lr} \left( 3 - 3 Lr + L^2 r^2 - \frac{\Lambda^2 r^2}{2} + \frac{Lr \Lambda^2 r^2}{2} \right)\\
    &+\text{erfc} \left( \frac{L}{\Lambda} - \frac{\Lambda r}{2} \right) e^{-Lr} \left( -3 - 3 Lr - L^2 r^2 + \frac{\Lambda^2 r^2}{2} + \frac{Lr \Lambda^2 r^2}{2} \right) \Bigg] + \frac{3e^{-\Lambda^2 r^2/4}}{4\pi^{3/2}\Lambda r^2},\\
	\nonumber
	r^2 \frac{\partial}{\partial r} \frac{1}{r} \frac{\partial}{\partial r} \frac{1}{r} \frac{\partial}{\partial r} s(r,y) = &\frac{e^{L^2/\Lambda^2}}{8\pi \Lambda^2 r^4} \Bigg[ \text{erfc} \left( \frac{L}{\Lambda} + \frac{\Lambda r}{2} \right) e^{Lr} \left( -10 + 10Lr - 5L^2r^2 + L^3r^3 - Lr\Lambda^2r^2 + \frac{L^2r^2\Lambda^2r^2}{2}  \right)\\
	\nonumber
    &+\text{erfc} \left( \frac{L}{\Lambda} - \frac{\Lambda r}{2} \right) e^{-Lr} \left( 10 + 10Lr + 5L^2r^2 + L^3r^3 - Lr\Lambda^2 r^2 - \frac{L^2r^2\Lambda^2 r^2}{2}  \right) \Bigg]\\
    &- \frac{e^{-\Lambda^2 r^2/4}( 10 + L^2r^2 + \Lambda^2 r^2 )}{4\pi^{3/2}\Lambda r^3},\\
	L(\vec q;y) = &\sqrt{m^2 + (1-y^2)\frac{\vec{q}^2}{4}}.
\end{align}
As a useful  cross-check,  we can see that in the limit of $q=0$ and $\Lambda \rightarrow \infty$   the above expression  reduces to
\begin{align}
    \frac{1}{8\pi r}\left[ (\vec{\sigma}_1\cdot\hat{\vec r})(\vec{\sigma}_2\cdot\hat{\vec r})(1+mr) - (\vec{\sigma}_1\cdot\vec{\sigma}_2) \right] e^{-mr}~,
\end{align}
which corresponds to Eqs. (5.8) and (5.9) in Ref.~\cite{Cirigliano2012}.
\end{appendices}
\twocolumngrid 

\bibliography{dm}

\end{document}